\documentclass[conference]{IEEEtran}
\usepackage{cite}
\usepackage{amsmath,amssymb,amsfonts}
\usepackage{algorithm}
\usepackage{algorithmic}
\usepackage{graphicx}
\usepackage{float}
\usepackage{textcomp}
\usepackage{xcolor}
\usepackage{caption}
\usepackage{graphicx}
\usepackage{float} 
\usepackage{subfigure}
\ifCLASSINFOpdf
\else
\fi
\begin{document}
%
\title{Secure and Scalable Circuit-based Protocol for Multi-Party Private Set Intersection}

\author{\IEEEauthorblockN{Jiuheng Su}
\IEEEauthorblockA{\textit{East China Normal University} \\
Shanghai, China \\
51215902053@stu.ecnu.edu.cn}
\and
\IEEEauthorblockN{Zhili Chen*}
\IEEEauthorblockA{\textit{East China Normal University} \\
Shanghai, China \\
zhlchen@sei.ecnu.edu.cn}
}


%


\IEEEoverridecommandlockouts
\makeatletter\def\@IEEEpubidpullup{6.5\baselineskip}\makeatother
\IEEEpubid{\parbox{\columnwidth}{
    Network and Distributed System Security (NDSS) Symposium 2024\\
    26 February - 1 March 2024, San Diego, CA, USA\\
    ISBN 1-891562-93-2\\
    https://dx.doi.org/10.14722/ndss.2024.23xxx\\
    www.ndss-symposium.org
}
\hspace{\columnsep}\makebox[\columnwidth]{}}

\maketitle

\begin{abstract}
We propose a novel protocol for computing a circuit which implements the multi-party private set intersection functionality (PSI). Circuit-based approach has advantages over using custom protocols to achieve this task, since many applications of PSI do not require the computation of the intersection itself, but rather specific functional computations over the items in the intersection.

Our protocol represents the pioneering circuit-based multi-party PSI protocol, which builds upon and optimizes the two-party SCS \cite{huang2012private} protocol. By using secure computation between two parties, our protocol sidesteps the complexities associated with multi-party interactions and demonstrates good scalability.

In order to mitigate the high overhead associated with circuit-based constructions, we have further enhanced our protocol by utilizing simple hashing scheme and permutation-based hash functions. These tricks have enabled us to minimize circuit size by employing bucketing techniques while simultaneously attaining noteworthy reductions in both computation and communication expenses.
\end{abstract}


%

\section{Introduction}
Two-party Private Set Intersection (PSI) enables two parties, denoted as $P_1$ and $P_2$ with respective input sets $X$ and $Y$, to compute the intersection $I = X \cap Y$ without revealing any other information about the items outside the intersection. Currently, there are numerous constructions of protocols for computing two-party PSI with concretely efficient and secure implementations. The problem of multi-party PSI (mPSI) naturally extends the concept of two-party PSI, i.e. $n$ parties collaborate to securely compute the intersection of their private input sets $S_1, S_2, ..., S_n$ while ensuring the confidentiality of all other information. 

Secure protocols for computing PSI, applicable to both two-party and multi-party scenarios, can be broadly categorized into two classes. The first category encompasses constructions specifically designed to address the PSI problem, yielding highly efficient protocols tailored to this particular task. On the other hand, the second category involves the utilization of generic secure multi-party computation (MPC) techniques, employing circuit representations of the desired functionality. This allows for the integration of PSI protocols into larger, composite secure computations. In this study, our focus is primarily directed towards the latter category of protocol constructions. These constructions offer the advantage of maintaining the secrecy of the intersection itself from the participating parties, while securely evaluating a symmetric function $f(S_1 \cap S_2 \cap ... \cap S_n)$, which could include operations such as set intersection sum or cardinality computation. For clarity, the functionality $F_{mPSI,f}$ is formally depicted in Figure 1.

\begin{table}[htbp]
\centering
\begin{tabular}{|l|}
\hline
\begin{tabular}[c]{@{}l@{}}
 \\
\textbf{Parameters} $P_1$ and $P_2$ are the two sides for computing the boolean \\ circuit $C$. $f$ is any symmetric function. \\ \\
\textbf{Input}: $P_i$ inputs set $S_i$, $i \in \{1, ..., n\}$.\\ 
\textbf{Output}: The functionality computes the circuit C on the parties' input \\ sets and returns the output $f(S_1 \cap S_2 \cap ... \cap S_n)$ to the parties. \\ \\

\end{tabular} \\ \hline
\end{tabular}
\captionof{figure}{Functionality $F_{mPSI,f}$}
\end{table}

In the context of two-party circuit-based PSI, we suppose that each party possesses an input set containing $n$ items. A most naive circuit requires $O(n^2)$ pairwise comparisons between the items. However, optimizations have been proposed by leveraging local computation capabilities of the parties. For instance, in the work of \cite{huang2012private}, several two-party circuit-based PSI protocols were introduced. For small universes, the parties can represent their input sets as bit-vectors and compute the intersection using bit-wise AND operations (referred to as the Bit-Wise And (BWA) protocol). On the other hand, for larger universes, \cite{huang2012private} presented the Sort-Compare-Shuffle (SCS) design. This design involves local sorting of the respective sets by each party, followed by the computation of the sorted list of the union of the two sets using the bitonic sorting network. Consequently, items in the intersection will appear twice adjacently in the sorted list, allowing for the identification of the intersection by comparing adjacent items. To preserve privacy, the sorted result of the intersection is then shuffled using a Waksman permutation network, effectively concealing the positional information of the items. The overall circuit computation requires only $O(n\log n)$ comparisons, primarily stemming from the initial stage of merge sort. The two-party circuit-based PSI protocols proposed by \cite{huang2012private} serve as the foundation for our research. In fact, the two-party circuit-based PSI protocols proposed by \cite{huang2012private} is the starting point of our work. 

\subsection{Overview of our Protocol}
Our protocols are built upon the foundation of the two-party circuit-based PSI protocols proposed by \cite{huang2012private}. One notable difference is that, in order to overcome the challenges of complex interactions and scalability in the multi-party setting, our protocol adopts a generic two-party secure computation protocol. As a result, prior to conducting circuit computations, the private inputs of the parties need to be securely distributed between the two parties engaged in the secure computation process. These two parties then reconstruct the items and perform secure computation of multi-party PSI (mPSI) within the circuit. The construction of our protocols uses generic secure multi-party computation (MPC) protocols, such as Yao's garbled-circuit protocol and the GMW protocol, to evaluate boolean circuits that compute the desired functionality. By relying on these established MPC protocols, which possess proven security properties, we can focus on designing circuits that effectively implement the desired functionality. In our study, we consider the multi-party setting involving $m$ parties denoted as $P_1, ..., P_m$. Each party possesses an input set consisting of $n$ items, which are represented using $\sigma$ bits.

In our first protocol, namely multi-party Bitwise-AND (mBWA), the input sets are represented as bit-vectors of length $2^{\sigma}$. The protocol incorporates a secret sharing scheme, where each party securely distributes their respective bit-vectors to two designated parties, denoted as $P_1$ and $P_2$. Subsequently, $P_1$ and $P_2$ reconstruct the bit-vectors within the circuit, enabling the computation of the intersection by performing bit-wise AND operations on the corresponding bit-vectors.

It is important to highlight that the practical applicability of this protocol is limited to scenarios involving smaller universes. The exponential growth of AND gates within the circuit imposes constraints on its scalability when dealing with larger datasets.

Our second protocol, multi-party Sort-Compare-Shuffle (mSCS), follows the Sort-Compare-Shuffle (SCS) paradigm while extending it to the multi-party setting. In this protocol, each party independently performs a local sorting operation on their respective input sets. The sorted sets are then distributed among the designated parties, $P_1$ and $P_2$, ensuring that the order of the secret shares aligns with the order of the items in the sorted sets.

Upon distribution, $P_1$ and $P_2$ reconstruct the sets within the circuit and merge them securely using a $k$-bitonic sorting network based on the $k$-Bitonic sort algorithm \cite{gao1999k}. The intersection of the sets can be subsequently determined by identifying the elements that occur $m$ times consecutively in the merged list. This identification process involves comparing adjacent elements. Finally, before revealing the sorted intersection, a shuffling procedure is applied to conceal the positional information of the items, thus preserving privacy and confidentiality.

Our research is also motivated by the work presented in \cite{pinkas2015phasing}, which introduced a method for comparing items mapped to each bin. In light of this, we propose a novel approach that employs simple hashing scheme for each party to distribute their elements into distinct bins. This approach aims to reduce communication overhead by utilizing multi-party circuit-based PSI protocols to compare the elements within each bin. Through our exploration, we have discovered substantial advantages in employing individual instances of the aforementioned protocol within each bin, as opposed to directly utilizing a single, large circuit for computation. These advantages extend beyond the realm of communication overhead and also encompass the facilitation of parallel computation. By leveraging this approach, our protocols exhibit improved efficiency and scalability, making them well-suited for practical deployment in multi-party settings.

\subsection{Motivation for multi-party Circuit PSI}
\subsubsection{Circuit}
Currently, the predominant focus of research efforts lies in addressing the PSI problem itself, which aims to reveal the intersection to the involved parties. These protocols have demonstrated high levels of efficiency, even achieving linear communication costs. However, in many practical applications, PSI functions as a module, and it is crucial to maintain the privacy of the intersection. In fact, these PSI applications often require the ability to compute any function based on the intersection.

Moreover, modifying or altering a custom MPC protocol has been shown to be prohibitively expensive and sometimes even infeasible. In contrast, generic MPC protocols offer greater flexibility in supporting additional computations through circuit expansion. They can leverage existing code bases and software packages, allowing users to focus on circuit design rather than developing an entirely new protocol. Clearly, the latter option is more challenging.

\subsubsection{Multi-party}
The multi-party PSI problem constitutes a more general case of the two-party PSI problem and presents greater potential in the context of massive data sharing. The multi-party scenario is better suited for various applications, such as identifying a target audience for an advertising campaign that involves several companies sharing data on their common users. It should be noted that generic Multi-Party Computation (MPC) protocols tend to be computationally expensive in the multi-party setting. Consequently, there exists a research gap in the context of multi-party circuit-based PSI. Nevertheless, as discussed earlier, there is significant motivation to address this gap and develop efficient solutions in this context.

\subsection{Related Work}
The problem of PSI has always been a hot issue in the field of MPC. We focus on the discussion of the state-of-the-art of semi-honest PSI protocols and simply classify previous works into two-party PSI and multi-party PSI. 
\subsubsection{Two-party PSI}
The earliest Private Set Intersection protocols were built upon public-key cryptography, specifically relying on the Diffie-Hellman assumptions, which can be traced back to the 1980s \cite{shamir1980power}. Subsequently, more efficient PSI protocols were developed based on oblivious transfer (OT) extension, which require minimal public-key cryptography computation and can be efficiently instantiated with symmetric key cryptography. Circuit-based PSI protocols use generic MPC protocols to perform the necessary computations.

A basic PSI circuit computes $O(n^2)$ element comparisons, resulting in $O(\sigma n^2)$ gates, where $\sigma$ represents the bit-length of the elements. The number of comparisons performed by the circuit is a crucial factor that impacts the overhead, as it directly affects the communication volume in the protocol. The Sort-Compare-Shuffle (SCS) PSI circuit introduced by \cite{huang2012private} reduces the number of element comparisons to $O(n\log n)$. The Circuit-Phasing PSI protocol proposed by \cite{pinkas2015phasing} hashes input items into $O(n)$ bins using Cuckoo hashing and simple hashing, enabling independent operations on each bin. Each bin typically contains at most $O(\log n/\log\log n)$ elements. Consequently, the Circuit-Phasing PSI circuit computes $O(n\log n/\log\log n)$ comparisons.

The first circuit-based PSI protocol to achieve linear communication complexity is presented in \cite{pinkas2019efficient}. This protocol relies on the use of oblivious programmable pseudo-random functions (OPPRF). Parties need to evaluate a circuit per bin to compare the programmed value with the output of the OPRF. As a result, this circuit only needs to compute one single comparison per bin.

\subsubsection{Multi-party PSI}
The first multi-party PSI protocol was introduced by \cite{freedman2004efficient}, which utilized oblivious polynomial evaluation (OPE) techniques like additively homomorphic encryption. Subsequent works by \cite{kissner2005privacy, sang2008privacy, cheon2012multi} focused on optimizing the computation and communication overhead of these protocols. The mPSI protocol proposed in \cite{kolesnikov2017practical} was the first implementation of multi-party PSI and introduced a novel primitive called Oblivious Programmable Pseudo-Random Functions (OPPRF). This protocol successfully avoids computationally expensive public-key operations.

To the best of our knowledge, the exploration of multi-party circuit-based PSI remains largely unexplored in the existing literature.

\subsection{Our Contributions}
In summary, in this paper we present the following contributions:

\begin{itemize}
\item We provide the first multi-party circuit-based PSI achieving $O(mnlog^2 (mn))$ asymptotic communication overhead. Our protocol is a natural generalization of the two-party circuit-based PSI protocol, which simplifies the complexity of multi-party interactions and can be easily expanded.

\item We integrate simple hashing scheme into our multi-party circuit-based protocols. By using the permutation-based hashing function, the elements can be represented in the form of shorter bits in the bins. Grouping data into bins results in a reduction in circuit size and enables parallel computation. This approach achieves the goal of simultaneously decreasing both communication and time overheads.
\end{itemize}

\section{Preliminaries}
\subsection{Setting}
 There are $m$ parties, which we denote as $P_1$, ..., $P_m$, where $P_1$ and $P_2$ are typically the two parties for security computation. Each of these parties is in possession of respective input sets, $S_1, S_2, \dots, S_m$, each of which contains $n$ items represented by $\sigma$ bits. It is assumed that $P_1$ and $P_2$ agree on a circuit $C$ that receives the secret shares of input sets and computes the intersection of $\hat{n}$ elements. They also agree on a symmetric function $f$ and can compute $f(S_1 \cap S_2 \cap ... \cap S_m)$ securely. We denote the computational and statistical security parameters by $\kappa$ and $\lambda$. We use $\gamma$ to denote a parameter that determines the probability of hashing failure, which is employed in optimization schemes. We use $S_i[j]$ to denote the $j$-th item in the set $S_i$. We also denote the set $\{1, ..., c\}$ as $[c]$. 
 
\subsection{Security Model}
In this work, similar to most protocols for private set intersection, we focus on the semi-honest model, also known as the honest-but-curious model, which assumes that all parties will follow the protocol, but adversaries may attempt to extract as much information as possible from the protocol execution. This is different from malicious adversary model, where adversaries can deviate from the protocol steps arbitrarily. While protocols designed for malicious adversaries offer more security, they tend to be less efficient than those designed for the semi-honest setting. In most scenarios, semi-honest security is sufficient, as it is currently difficult for adversaries to modify software with attestation or business restrictions. However, for most recent optimization of circuit-based PSI protocols that rely on Cuckoo hashing, it is still difficult to ensure that such operation of hashing is secure and correct. 

The SCS protocol of \cite{huang2012private} is a unique circuit-based PSI protocol that can be easily modified to provide security against malicious adversaries by expanding the circuit to verify that the elements are sorted, while maintaining an overall complexity of $O(n log n)$. It is worth noting that this advantage is also present in our protocol.

\subsection{Secure Two-Party Computation}
In contemporary MPC research, there exist two primary methods for the secure computation of boolean circuits: Yao's garbled circuit (GC) protocol \cite{yao1986generate} and the GMW protocol \cite{goldwasser1987play}. 

Yao's garbled circuit protocol presents a constant round complexity and implements the free XOR gates technique \cite{kolesnikov2008improved}. Through the optimization techniques developed in \cite{zahur2015two}, the protocol requires at least two ciphertext transmissions to evaluate an AND gate. Similarly, the GMW protocol also implements the free XOR technique and necessitates two ciphertext transmissions to evaluate each AND gate using OT extension \cite{asharov2013more}. However, the GMW protocol offers an additional benefit in the form of its ability to perform symmetric cryptographic computations in advance during the pre-computation phase, thereby improving the efficiency of the online phase. 

The main advantage of generic protocols is that it can easily extend the functionality of the protocol without having to change the security of the protocol. As such, we use generic secure two-party computation protocol to implement our protocols.  

\subsection{Secret Sharing}
In cryptography, an $(n, t)$-secret sharing scheme has been proposed for distributing a secret $s$ among $n$ parties in such a way that any $t + 1$ parties can collectively reconstruct the secret $s$ from their shares, while preventing any collusion of $t$ parties from learning any information about $s$ \cite{shamir1979share, blakley1979safeguarding}. This scheme provides a secure and efficient way to distribute secret information among multiple parties without compromising its confidentiality. 

In this context, our protocols employ an additive $(n, n-1)$-secret sharing scheme. In our protocols, all parties have to distribute their input sets among two designated parties, $P_1$ and $P_2$. This distribution ensures that neither $P_1$ nor $P_2$ can obtain any information about the input sets of other parties except their own data. During the computation phase of the circuit, $P_1$ and $P_2$ reconstruct the inputs of the parties in the circuit and calculate the intersection of the input sets.

The use of the additive secret sharing scheme in our protocols provides an additional layer of security to the distribution of secret information among multiple parties. The data pre-processing phase ensures that the input sets of parties are kept confidential, and the computation phase guarantees that the secret information is reconstructed securely without revealing any information to unauthorized parties. This approach is beneficial for applications that require the distribution of confidential information among multiple parties, such as secure multi-party computation, privacy-preserving data analysis, and secure cloud computing. 

\subsection{Simple Hashing}
We have incorporated a hashing scheme in our protocols to optimize them. The literature on hashing schemes is extensive and covers a range of methods for handling collisions, complexities associated with insertion, deletion, and look-up operations, as well as utilization of storage space. Previous works such as \cite{pinkas2014faster, pinkas2015phasing, freedman2016efficient} have used hashing to improve the number of comparisons performed in Private Set Intersection (PSI) protocols. Similarly, our protocols allow the use of simple hashing schemes to split the computation by mapping input items to bins. 

The simple hashing scheme typically utilizes a table $T$ containing $\beta$ bins. We make the assumption that the number of bins $\beta$ is a power of 2. In cases where $\beta$ is not a power of 2, \cite{arbitman2010backyard} proposes a method to handle this situation. Using a hash function $H$ which maps an element $e$ to an address $a = H(e)$ within the range $[0, \beta-1]$, the element $e$ is then placed into the corresponding bin $T[a]$. The simple hashing approach allows for multiple elements to be stored in each bin, with the maximum number of elements that can be stored in each bin depending on the total number of elements and the number of bins. This problem has been analyzed in detail in \cite{raab1998balls}, which showed that when randomly mapping $n$ items to $n$ bins using $H$, the most populated bin would have at most $\frac{ln n}{ln ln n}(1 + o(1))$ items with high probability. 

In summary, by adopting simple hashing scheme, our protocols allow for efficient computation by reducing the number of comparisons while enabling parallel computation.

\subsection{Permutation-based Hashing}
When dividing elements into bins, it is possible to reduce the bit-length of stored items through permutation-based hashing, a technique introduced in prior literature \cite{arbitman2010backyard, pinkas2015phasing}. In this work, we apply permutation-based hashing to improve the memory usage of our simple hashing scheme. Specifically, by utilizing permutation-based hashing, the elements stored in bins can be represented using fewer bits, resulting in a reduction of the number of gates required during the circuit computation stage. This reduction can lead to significant efficiency improvements in terms of communication costs and computation time. Notably, the permutation-based hashing technique can be applied in all hashing-based privacy-preserving set intersection (PSI) protocols, including the one proposed in this study.

In general, a traditional hash function $h:\{0, 1\}^{\sigma} \rightarrow \{0, 1\}^{log \beta}$ maps an element $x$ to bin $h(x)$, where $|x|$ represents the bit-length of $x$. We notice that the bin index $h(x)$ is able to carry $log \beta$ bits of information. So, it is possible to reduce the information stored in the bins from $\sigma$ to $\sigma - log \beta$, which means that secure computations are done on elements with smaller representations. If we choose the hash function carefully, we can realize that if two items have the same representation stored in the bin and are in the same bin, they must be equal. Permutation-based hashing provides uses a Feistel-style trick to implement this purpose. 

In details, we split the bit representation of an input item $x$ into $x_L || x_R$, where $x_L$ has $log \beta$ bit-length, which is equal to the big-length of the bin index in the hash table, and $|x_R|$ has $\sigma - log \beta$ bit-length. Then we define $f ()$ be a random function whose range is $[0, \beta - 1]$, represented by $log \beta$ bits. We define $h(x) = x_L \oplus f(x_R)$. Then the input item $x$ is stored in the bin $x_L \oplus f(x_R)$, and the value stored in the bin is $x_R$, which is reduced to $\sigma - log \beta$ bit-length. We observe that if two elements $x$ and $y$ are stored in the same bin, and the stored values $x_R$ and $y_R$ are also the same, then that means $f(x_R) = f(y_R)$. Since the bin $h(x) = h(y)$, then $x_L = y_L$. So, we can conclude that $x = y$. Note that if $|x|$ is not much longer than $log \beta$, the overhead will have a great improvement. 

\begin{table*}[htbp]
\centering
\begin{tabular}{|l|}
\hline
\begin{tabular}[c]{@{}l@{}}
\\
$P_1$ and $P_2$ are the two sides for computing the boolean circuit $C$.  \\ \\
\textbf{Input to} $P_i$: Input set $S_i$, $i \in [m]$. \\ 
\textbf{Output}: The intersection $I$ of the input sets. \\ \\

\textbf{Preparation}: \\
\hspace*{0.1in}--- $P_i$ represents the input set $S_i$ as bit-vector $v_i$ of length $2^{\sigma}$ locally. \\ 
\hspace*{0.1in}--- $P_i$ randomly selects $r_i$ and $r_i'$ so that $r_i + r_i' = v_i$. \\ \\

\textbf{Execution}: \\ 
\hspace*{0.1in}1. $P_i$ sends $r_i$ to $P_1$ and $r_i'$ to $P_2$, for all $i$, we have $r_i + r_i' = v_i$. \\ 
\hspace*{0.1in}2. $P_1$ and $P_2$ input the received secret share, $\{r_i\}_{i \in [m]}$ and $\{r_i'\}_{i \in [m]}$, into the boolean circuit $C$ and perform secure two-party computation: \\
\hspace*{0.3in} --- Reconstruct the bit-vectors $\{v_i\}_{i \in [m]}$ using $m2^{\sigma}$ XOR gates. \\
\hspace*{0.3in} --- By performing AND operations on bit vectors $\{v_i\}_{i \in [m]}$ in turn, the circuit C outputs the ciphertext of the bit-vector representation of the \\ \hspace*{0.45in} intersection. \\
\hspace*{0.1in}3. $P_1$ and $P_2$ send the result (the bit-vector of the intersection) to all the other parties. \\
\hspace*{0.1in}4. All participants infer the elements of the intersection according to the representation of the bit-vector of the intersection. \\ \\

\end{tabular} \\ \hline
\end{tabular}
\captionof{figure}{Multi-party Bitwise-AND Protocol}
\end{table*}

\subsection{Circuit PSI based on Hashing}
The first circuit-based PSI protocol was introduced in \cite{huang2012private}. Prior to performing computations in the circuit, the parties are required to sort their input sets and input them to the circuit. In the protocol proposed in \cite{pinkas2015phasing}, the parties must also utilize hash schemes, such as Cuckoo hashing and simple hashing, to map their input items into bins. Generally, these circuit-based PSI protocols necessitate that the parties conduct operations (e.g., reordering and hashing) on their input sets beforehand.

It was shown in \cite{freedman2004efficient} that if the parties map their input items into bins, then they only need to compare the items that stored in the same bins. Nevertheless, the number of items mapped into each bin may reveal information about their input sets. Thus, to ensure that this information is concealed from other parties, it is necessary to pad all bins with random dummy values, without revealing the number of items in each bin. It is assumed that the parties agree on the maximum number of items that can be mapped into a bin, denoted as $B$. Following the mapping of all items into bins, the parties must pad each bin with dummy values until it contains $B$ items. If the two parties compare the items in the bins using pairwise comparisons, the total number of the comparisons will reduce from $O(n^2)$ to $O(\beta \cdot B^2)$, where $n$ is the number of items in each input set and $\beta$ is the number of the bins. 

In our proposed protocols, we combine simple hashing scheme with the Sort-Compare-Shuffle protocol to compare items in each bin. By carefully selecting the number of bins $\beta$ and the upper bound $B$ for the maximum number of items that can be mapped to a bin, this approach can offer computational and communicational advantages over previous schemes. In this scenario, the parties need to map their input items into bins using a hash function and locally sort the items in each bin for subsequent merge sorting.

\section{Multi-party Bitwise-AND Protocol}
As illustrated in Figure 2, the mBWA protocol is exclusively viable for small-scale universes. In such circumstances, the input sets can be succinctly represented as bit-vectors of length $2\sigma$. In a two-party context, the intersection can be computed by applying a simple bit-wise AND operation between the bit-vectors of the two parties. When it comes to multi-party situation, each party must first distribute their respective input sets among two designated parties (typically denoted as $P_1$ and $P_2$) through the use of a secret-sharing scheme prior to the computation phase. Subsequently, $P_1$ and $P_2$ must reconstruct the respective bit-vectors of each party. The intersection can then be computed by performing a bit-wise AND operation between the bit-vectors.

The whole computation process of the circuit is relatively clear and the circuit can be obtained by instantiating a binary XOR gate $m2^{\sigma}$ times and a binary AND gate $(m - 1)2^{\sigma}$ times. The utilization of the free XOR gates technique permits the XOR gates in the circuit to be evaluated without incurring any communication or cryptographic operations, resulting in a bit-vectors reconstruction stage that is free of such operations. Despite the exponential number of AND gates, the small constant factor leads to favorable performance, particularly after integrating the optimization of the hash scheme. Detailed information about the experimental results can be found in the "Experimental Results" section.

\section{Multi-party Sort-Compare-Shuffle Protocol}
It is worth noting that while the cost of a multi-party protocol increases exponentially with sigma, the constant factor is relatively small, leading to satisfactory performance in small universes. However, for larger universes, it is necessary to further reduce the protocol overhead. To this end, we propose the mSCS protocol, following the SCS paradigm. The protocol leverages the local computing capabilities of each party to minimize overhead.

\begin{figure}[htbp]
\centerline{\includegraphics[width=3.5in]{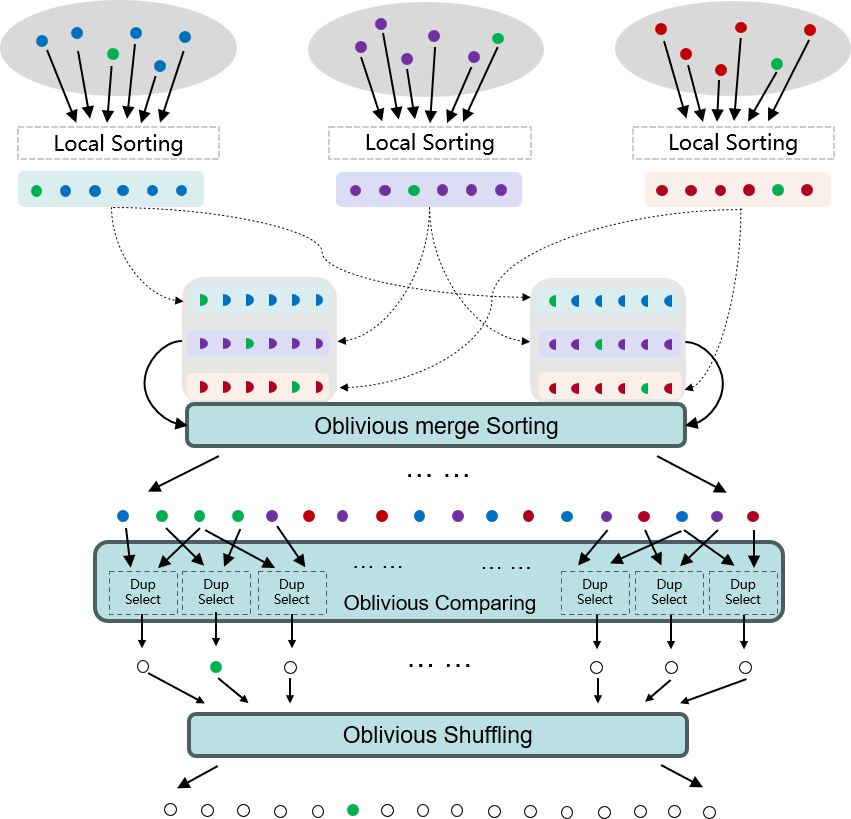}}
\caption{Example of multiparty Sort-Compare-Shuffle protocol where $m = 3$ (any stage labeled as "Oblivious" in the figure indicates the parts that require two-party secure computation).}
\label{fig}
\end{figure}

The mSCS protocol comprises three parts as shown in Figure 3. Firstly, each participant, including $P_1$ and $P_2$, performs local sorting on the input set and then distributes the sorted set to $P_1$ and $P_2$ via a additive secret sharing scheme. Secondly, $P_1$ and $P_2$ utilize generic secure two-party computation to reconstruct the input sets and sort the union of the two sets with an oblivious merging network. As the sequence is already sorted, it is not necessary to compare every adjacent element, as was done in the scenario involving two parties. Rather, comparison of elements at specified positions in the sorted sequence suffices for finding the intersection elements. For example, in a scenario with three participating parties, if the first and third elements are equal, then the element must be one of the elements in the intersection. Nonetheless, direct output of the intersection elements is not viable, as this may reveal positional information. Further details can be found in \cite{huang2012private}. Therefore, it is imperative to shuffle the matched elements to conceal any positional information from the resulting order.

\subsection{Sort}
Referring to the first protocol for distributing and reconstructing the input sets, we assume that the input set of each participant has already been reconstructed in the circuit. Following this, participants $P_1$ and $P_2$ are required to implement an oblivious merging network to sort the union of the input sets, making use of the fact that the input sets are already sorted. The term "oblivious" implies that regardless of the order of the input elements, the circuit remains fixed, i.e., the sequence of comparison between elements is predetermined. In order to merge the $m$ sorted lists into a fully sorted sequence, a merge-sort network is designed based on the $k$-Bitonic sort algorithm \cite{gao1999k}. The resulting number of comparisons is $O(mnlog^2 (mn))$.

\begin{figure*}[htbp]
\centering
\subfigure[2-Sorter]{
\begin{minipage}[t]{0.3\linewidth}
\centering
\includegraphics[width=1.9in]{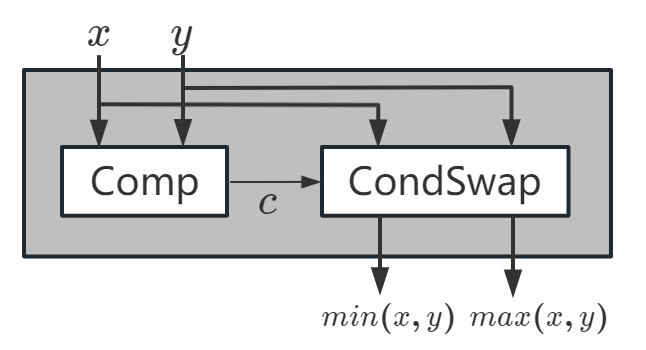}
\end{minipage}
}
\subfigure[1-bit Comparator]{
\begin{minipage}[t]{0.3\linewidth}
\centering
\includegraphics[width=2.2in]{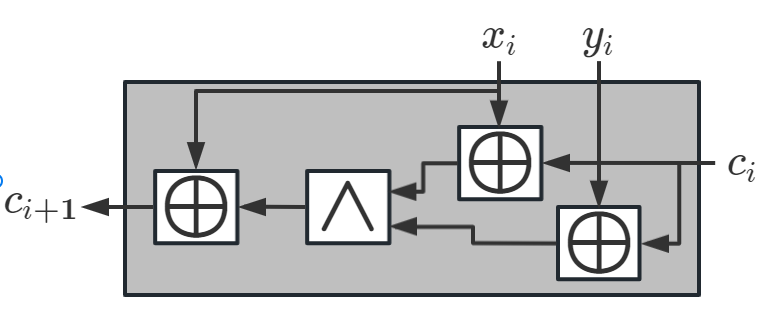}
\end{minipage}
}
\subfigure[1-bit CondSwap]{
\begin{minipage}[t]{0.3\linewidth}
\centering
\includegraphics[width=1.4in]{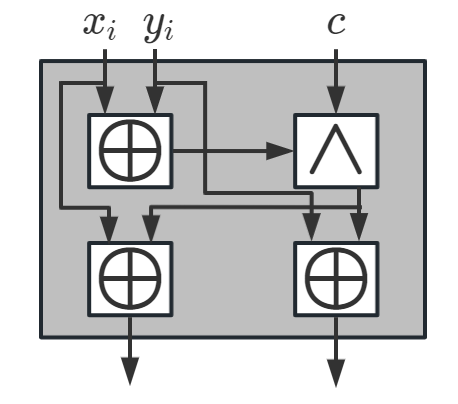}
\end{minipage}
}
\centering
\caption{Basic modules of a sorting network.}
\end{figure*}

The concatenation of a sequence sorted in ascending order with a sequence sorted in descending order yields a bitonic sequence, which can be obtained after local sorting. In the context of sorting networks, the 2-Sorter module serves as a fundamental component. As shown in Figure 4, a design for 2-Sorter is presented in \cite{huang2012private}, which comprises a $\sigma$-bits Comparator and a $\sigma$-bits CondSwap, resulting in a requirement of only $2\sigma$ non-free gates to compare two $\sigma$-bits elements. We can recursively utilize the Batcher's bitonic sorting network, as described \cite{huang2012private}, in a tree-like manner. However, the resulting high complexity caused by excessive redundant computations is not acceptable.

In our work, we introduce a $k$-Bitonic sort algorithm, denoted as Algorithm 1, which is a generalized version of the bitonic sort introduced by Batcher in 1968. We assume that $k = \lceil \frac{m}{2} \rceil$ to indicate that there are $k$ bitonic sequences in total in the case of $m$ parties. It is important to note that $k$-Bitonic sort reduces to Batcher's bitonic sort when $k$ is equal to 1. Specifically, the code executed until the 10th line of the algorithm is identical to Batcher's bitonic sorting algorithm. 

We assume that the input $V = V[0:N - 1]$ is a $k$-bitonic sequence, where $N = mn$ is the length of the sequence, and the ouput $U[0:N-1]$ is an incremental sequence. By performing odd-even splitting on a sequence, the resulting sub-sequences can still exhibit the characteristic of $k$-bitonic sequence. Thus, the $k$-bitonic sort algorithm KBS($V,N,U$) is a recursive procedure. We denote $Q({a_0, a_1, ..., a_{n-1}})$ is the permutation of the sequence $A = \{a_0, a_1, ..., a_{n - 1}\}$ ordered in an ascending fashion. The fundamental operation in the execution process of the KBS algorithm is to compare and exchange two elements. 

\begin{algorithm}[H]
\caption{\textit{$k$-Bitonic Sort KBS($V, N, U$)}}
\begin{algorithmic}[1]
\IF{N == 1}
\STATE RETURN;
\ENDIF
\STATE $Y[0:\frac{N}{2} - 1, 0:1] = V[0:N-1]$; ($Y[i, j] = V[2i + j]$)
\FOR{$j = 0$ to $1$}
\STATE \textit{KBS}($Y[*, j], \frac{N}{2}, B[*, j]$);
\ENDFOR
\FOR{$i = 0$ to $\frac{N}{2} - 1$}
\STATE $C_0[i, *] = Q(B[i, *])$; 
\ENDFOR
\STATE $d = \lceil log_2(N) \rceil$
\IF{$N \leq 2k$}
\STATE $d = d - 1$;
\ENDIF
\FOR{$t = 1$ to $d$}
\STATE $\delta = 2^{d-t}$;
\FOR{$i = 0$ to $\frac{N}{2} - 1 - \delta$}
\STATE $(C_t[i, 1], C_t[i+\delta, 0]) = Q(C_{t-1}[i, 1],C_{t-1}[i+\delta, 0])$;
\ENDFOR
\ENDFOR
\STATE $U = C = C_d$;  ($U[2i + j] = C[i,j]$)
\end{algorithmic}
\end{algorithm}

 Based on the complexity analysis of the recursive function, we deduce that the time complexity of the KBS algorithm is $O(mnlog^2(mn))$. To provide a more detailed analysis, we use the recurrence formula to work out the number of comparisons performed by the merge sort circuit. The resulting expression is $\frac{mn}{4}log(mn)log(\frac{mn}{2}) + mn - 1$. Thus, we can construct a circuit that merges $m$ sequences of $n$ elements, each consisting of $\sigma$ bits, into a single sorted sequence of $mn$ elements. The circuit requires $2\sigma(\frac{mn}{4}log(mn)log(\frac{mn}{2}) + mn - 1)$ non-free gates.

\subsection{Compare}
After sorting the $mn$ elements in a list, the next step in the secure set intersection protocol involves comparing adjacent $m$ elements to identify the elements in the intersection.

In this phase, we employ a duplicate-selection circuit to identify all elements in the intersection.  Specifically, the circuit filters out the elements in the intersection by computing whether a consecutive set of $m$ elements are all equal. If they are equal, it outputs the value of the element; otherwise, it outputs a dummy value. Our investigation focused on exploring the properties of sorted lists and their relevance to the protocol at hand. We aimed to gain a deeper understanding of the characteristics that could be leveraged to optimize the execution of the protocol. After thorough analysis, we identified two key properties that are particularly significant: 

\subsubsection{Non-adjacent element comparison} By exploiting the inherent order of the sorted list, we discovered that it is unnecessary to compare adjacent elements during the computation. Instead, we can employ a larger stride when comparing elements. In a scenario involving three parties, we denote the the sorted list as $l$. So we can compare elements $(l_i, l_{i + 2})$ rather than evaluating pairs $(l_i, l_{i + 1})$ and $(l_{i + 1},l_{i + 2})$. This strategic adjustment effectively reduces the computational complexity of the circuit, leading to improved efficiency.

\begin{figure}[htbp]
\centerline{\includegraphics[width=3.2in]{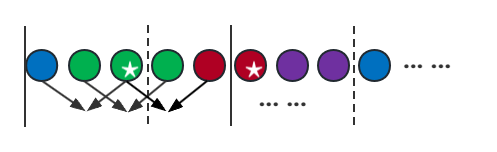}}
\caption{Example of comparing three consecutive elements.}
\label{fig}
\end{figure}

\subsubsection{Obliviousness of matched elements} Through our investigation, we observed an intriguing pattern within the sorted sequence. In any consecutive group of $2m-1$ elements, there can be at most one matched element, and it will always be positioned in the middle of the group. This critical characteristic aligns with the requirement of the circuit's obliviousness, ensuring that only relevant and meaningful elements are considered in the intersection computation. 

Based on these two properties, we are able to streamline the circuit design and enhance its overall performance. To implement and leverage the two important properties we have discovered, as shown in Figure 6, we design a $5$-duplicate-selection circuit to acquire the matched elements in the intersection, if any. Specifically, the circuit takes as input $5$ consecutive elements and outputs the matched element or a dummy value $0^{\sigma}$. In fact, the combination of the $5$-duplicate-selection circuits ensures that every continuous $5$ elements of the sequence in the circuit will be compared. 

\begin{figure*}[htbp]
\centering
\subfigure[5-duplicate-selection circuit]{
\begin{minipage}[t]{0.3\linewidth}
\centering
\includegraphics[width=2.1in]{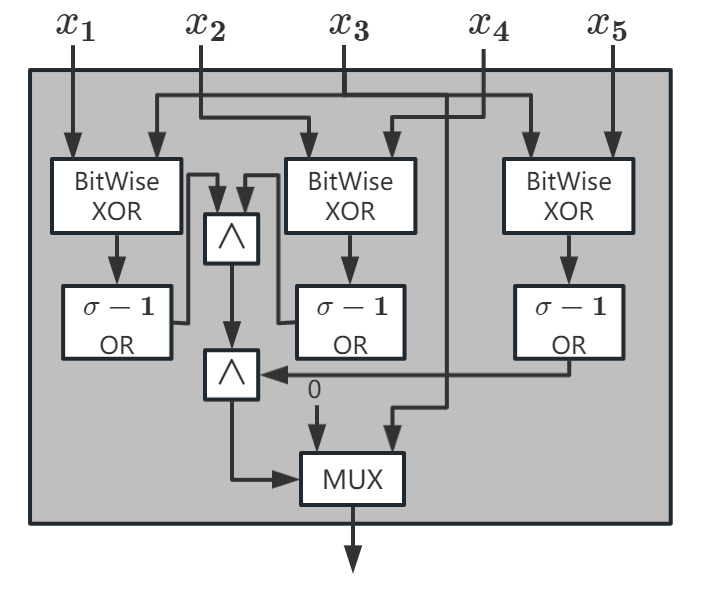}
\end{minipage}
}
\subfigure[Combination of 5-duplicate-selction circuits]{
\begin{minipage}[t]{0.6\linewidth}
\centering
\includegraphics[width=4.5in]{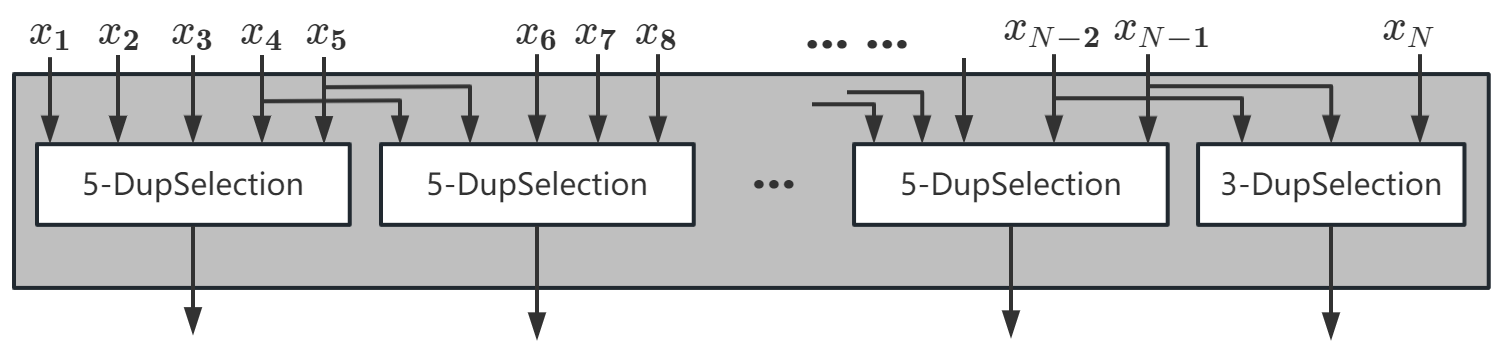}
\end{minipage}
}
\centering
\caption{Design and use of 5-duplicate-selection circuits.}
\end{figure*}

The $5$-duplicate-selection circuit is constructed using $4\sigma - 1$ non-free gates. The combination of duplicate-selection circuits employs $n - 1$ $5$-duplication-selection circuits and one $3$-duplication-selection circuit which can be constructed using $2\sigma - 1$ non-free gates. Consequently, the total number of non-free gates required for this phase is $(4n - 2)\sigma - n$. By leveraging the property of non-adjacent element comparison, in a scenario involving three participants, we can achieve a reduction in the circuit's size by approximately 25\% for this particular stage. Furthermore, the effectiveness of this optimization becomes more pronounced when $m$ is a relatively large value (more than 10). Moreover, this particular circuit design is capable of producing $n$ elements rather than $3n - 2$ which would be produced using adjacent element comparison. Therefore, this optimization also results in a reduction of approximately 60\% in the size of the circuit for the next stage (Shuffle).

This design can be generalized to $m$ participants, where the total number of non-free gates required for the comparison phase is $[(m + 1)\sigma - 1](n - 1) + 2\sigma - 1$. Therefore, we have devised and employed a combination of duplicate-selection circuits, which compare each of the $m$ consecutive elements in the sorted sequence, in order to figure out the elements in the intersection. This process involves the use of $O(mn)$ non-free binary gates. 

\subsection{Shuffle}
Upon figuring out the intersection, the matched $n$ elements (and dummy values) are arranged in a specific positional order in the circuit. This ordering has the potential to compromise the confidentiality of parties' input sets. However, if we only need to perform subsequent computations based on the elements in the intersection, this step can be omitted.   Therefore, to preserve privacy, it is essential to shuffle the elements before their disclosure. The shuffling process is crucial to ensuring privacy in secure multi-party computation, especially when dealing with sensitive data. In the absence of a proper shuffling algorithm, the positional information of the elements in the circuit may allow an adversary to deduce the corresponding parties' input sets. Specific examples can be obtained in \cite{huang2012private}. In line with \cite{huang2012private}, we also utilize an oblivious shuffling network, which is constructed using $O(mnlog(mn))$ non-free gates, to implement the random permutation needed to obliterate the positional information.

An oblivious shuffling network employs a set of gates that do not reveal any information about the input values while transforming the input sequence into an output sequence. The network achieves this by iteratively swapping pairs of elements in the sequence according to a predetermined permutation. Since the permutation is randomly generated, the resulting sequence is uniformly distributed over all possible permutations. Consequently, the positional order of the elements in the circuit is destroyed, and parties' input sets remain private.

The computational complexity of the shuffling process is a critical consideration, as it determines the scalability of the multi-party computation protocol. The $O(mnlog(mn))$ complexity of the oblivious shuffling network used in this protocol can be expensive for large inputs. Nonetheless, it remains a practical solution for most use cases, and ongoing research aims to develop more efficient shuffling algorithms.

\section{Hashing to Bins}
In this section, we present an enhanced iteration of the previously proposed protocol. Our approach combines the simple hashing scheme with circuit-based multi-party private set intersection to minimize communication overhead and enable parallel computing. We consider the earlier proposed protocol as a sub-protocol within our optimization framework. Specifically, we employ the simple hashing scheme to partition the data into multiple bins, and subsequently employ the circuit-based sub-protocol within each bin to compute the set intersection efficiently. This optimized approach aims to improve the overall performance and scalability of the protocol while maintaining the privacy guarantees provided by the original scheme.

To further improve the efficiency of the protocol, we also use the permutation-based hashing function. These hash functions allow for a more compact representation of the elements stored in each bin, which can further reduce communication costs, as the communication cost mainly depends on the number of non-free gates in the circuit. 

We provide a detailed description of the main ideas behind our optimization approach and the overall protocol flow. To prove the effectiveness of our optimization approach, we also conduct extensive calculations and analyses on both the efficiency and security of our protocol after incorporating the hashing scheme.

\subsection{Construction}
As shown in Figure 7, each participant follows simple hashing scheme, where each bin can store more than one element, using a public permutation-based hash function $h$ to hash elements from the input set to their respective bins. We assume that each participant possesses $\beta$ bins, with each bin having a capacity of $B$, including both dummy values and actual elements. 

Once the elements are hashed into the bins, the generic circuit-based multi-party protocol is executed in each corresponding bin to perform the set intersection operation. The circuits in each bin are independent of each other, allowing for simple and efficient parallel computation.

\begin{table*}[htbp]
\centering
\begin{tabular}{|l|}
\hline
\begin{tabular}[c]{@{}l@{}}
\\
$P_1$ and $P_2$ are the two sides for computing the boolean circuit $C$.  \\ \\
\textbf{Input to} $P_i$: Input set $S_i$, $i \in [m]$. \\ 
\textbf{Output}: The intersection $I$ of the input sets. \\ \\

\textbf{Preparation}: \\
\hspace*{0.1in}--- Let $h: \{0, 1\}^{\sigma} \rightarrow \{0, 1\}^{log \beta}$ be a public hash function. \\ \\

\textbf{Execution}: \\ 
\hspace*{0.1in}1. $P_i$ applies the hash function $h$ to hash all of their elements into respective $\beta$ bins. If the number of elements in any bin exceeds the capacity $B$, \\
\hspace*{0.18in} the protocol is terminated. \\ 
\hspace*{0.1in}2. For each bin, $P_1$ and $P_2$ perform secure computation on each bin to implement circuit-based multi-party private set intersection protocol (mBWA \\ 
\hspace*{0.18in} and mSCS).  \\
\\

\end{tabular} \\ \hline
\end{tabular}
\captionof{figure}{Combining simple hashing and circuit-based multi-party private set intersection protocol}
\end{table*}

\subsection{Security}
In our proposed scheme, each party employs simple hash scheme to map their respective input set into multiple bins. In simple hash scheme, if the number of items mapped to a bin exceeds its capacity, the use of bins with a constant size may result in hashing failures. 

When hashing fails, the party responsible for the hashing operation has two possible options. The first option is to ignore the unmapped element and remove it from its input set, which may result in an incorrect final computed result, albeit rare. The second option is to attempt to use an alternative hash function to remap the unmapped element. However, this approach requires informing the other parties involved of the use of the new hash function, which introduces a potential privacy leak. For instance, the other party could infer whether the input set $S$ of the first party could be equal to a set $S'$ by checking if $S'$ did not encounter hash failure, indicating that $S'$ and $S$ are not identical. Thus, it is essential to carefully set the capacity of each bin to minimize the probability of hashing failures to be negligibly small. However, a large bin capacity may inevitably result in an excessive number of dummy values in each bucket, leading to a reduction in the overall efficiency of the protocol. 

The most desirable scenario is when all elements are uniformly mapped to bins and each bin is fully occupied. In this case, we denote the ideal capacity of each bin is $b = \frac{n}{\beta}$, where $n$ is the total number of elements and $\beta$ is the number of bins. Next, the actual capacity $B$ should be as close as possible to $b$ while maintaining a negligible probability of hashing failures. We assume that $h$ is a random uniform hash function, and $X_i, i \in [n]$ are independent random variables. From the perspective of a fixed party, observing a fixed bin, $X_i= 1$ means that the $i$-th element is mapped to the bin by $h$, otherwise $X_i = 0$. Then, $X=\sum_{i=1}^nX_i$ represents the number of the participant's elements mapped to that bin. Thus, we can obtain $E(X) = b$.

In probability theory, \textit{Chernoff Bound} is a statistical concept that helps us understand the probability of the sum of independent random variables deviating from its expected value. Thus, according to Chernoff Bound, we can obtain that for any $\delta \in [0,1]$, it holds that:
$$Pr[X \geq (1 + \delta)b] \leq e^{-\frac{\delta^2b}{3}}$$

We define the event $A_i$ as $X \geq (1+\delta)b$, which represents the occurrence of hashing failure in a particular bin. Using the Boolean inequality $P(\bigcup_i A_i) \leq \sum_i P(A_i)$, given that there are $m$ parties and each party has $\beta$ bins, the probability of the whole protocol experiencing hashing failure satisfies: 
\begin{align}
    \begin{aligned}
        Pr[Failure] &\leq m \beta e^{-\frac{\delta^2b}{3}} \\
             &= m \frac{n}{b} e^{-\frac{\delta^2b}{3}} \\
             &= 2^{log_2e^{-\frac{\delta^2b}{3}} + log_2\frac{mn}{b}} \\
             &< 2^{-\frac{\delta^2b}{3} + logn} \nonumber
    \end{aligned}
\end{align}

Therefore, if $B$ is set to $(1 + \delta)b$ and the right-hand side of the inequality is a negligible function, it suffices to prove that the proposed scheme will result in negligible probability of hashing failures. So, if we want to keep the overall failure probability less than $2^{-\gamma}$: 
\begin{align}
    \begin{aligned}
        2^{-\frac{\delta^2b}{3} + logn} &< 2^{-\gamma} \nonumber
    \end{aligned}
\end{align}
We can get:
$$b > \frac{3(logn + \gamma)}{\delta^2}$$
and
$$\beta = \frac{n}{b} < \frac{n\delta^2}{3(logn + \gamma)}$$ 

\begin{table*}[htbp]
  \centering
    \begin{tabular}{|c|c|c|c|c|c|c|c|c|c|c|c|c|}
    \hline
    m & \multicolumn{3}{|c|}{3} & \multicolumn{3}{|c|}{5} & \multicolumn{3}{|c|}{7} & \multicolumn{3}{|c|}{9} \\
    \hline
    n & $2^8$ & $2^{12}$ & $2^{16}$ & $2^8$ & $2^{12}$ & $2^{16}$ & $2^8$ & $2^{12}$ & $2^{16}$ & $2^8$ & $2^{12}$ & $2^{16}$ \\ 
    \hline
    Time & 0.37 & 7.07 & 160.39 & 0.52 & 12.30 & 272.80 & 0.73 & 19.78 & 357.26 & 0.97 & 32.90 & 593.54 \\
    \hline
    Comm. & 13.68 & 430.46 & 11457.65 & 48.97 & 1539.40 & 41018.387 & 100.14 & 3151.91 & 83869.97 & 165.53 & 5208.56 & 138637.57 \\ 
    \hline
    \end{tabular}
    \caption{Total runtime (in seconds) and communication (in MB) of Plain mSC protocol. All the parties have n 32-bit elements as input.}
\end{table*}

\subsection{Complexity Analysis}
In fact, we can set $\delta = 1$ and $b = \log^2n$ to make the probability of overall hashing failure a negligible function of the input set size $n$. If we consider the mSCS protocol and use it to handle the elements in each bin, we need to perform $O(mb \log^2(mb))$ comparisons between elements. Thus, the total number of comparisons for $\beta$ bins is $O(\beta mb \log^2(mb))$, which can be simplified as $O(mn log^2(mlogn))$.

We assume that all the elements can be represented using $\sigma$ bits. By using permutation-based hashing function to map elements to corresponding bins, the elements stored in the bin can be represented with shorter bits. Specifically, with $\beta$ bins, the corresponding element can be stored using $l = \sigma - log\beta$ bits, and it can be guaranteed that if two elements have the same value stored in the same bin, then these two elements must be equal. Therefore, the asymptotic communication complexity of our final protocol reduces from $O(\sigma mn log^2(mn))$ to $O(l mn log^2(mlogn))$. By summing up, simplifying, and scaling down the number of non-free gates across the three stages, we obtain an upper bound on the number of non-free gates required for computations within each bin:  
\begin{alignat}{2}
    & \quad \sigma[\frac{mn}{2}log^2(mn) + \frac{8mn}{3} + n] \nonumber
\end{alignat}
In our proposed optimization scheme, each bin stores $B = (1 + \delta)b$ elements, and each element has a length of $\sigma - \log\beta$ bits. Therefore, we can obtain an upper bound on the number of non-free gates for each bin required after incorporating the hashing scheme using the above equation (with $n = (1 + \delta)b$ and $\sigma$ is set to $\sigma - \log\beta$). That is, for given $m$, $n$, $\sigma$, and $\gamma$, we can minimize the number of non-free gates required by setting the value of $\delta$ and $b$.

\section{Experimental Results}
In this section, we provide a comprehensive evaluation of the performance and costs associated with our protocols, considering specific values for the security parameters. Specifically, we set the computational security parameter to $\kappa = 128$, and the statistical security parameter to $\lambda = 80$. As the work of \cite{huang2012private} has demonstrated, we find that the BWA protocol is the optimal choice when the size of the element space is small (up to approximately $\sigma$ = 20), as verified through experimental validation. Therefore, to accommodate more general scenarios, we have exclusively implemented mSCS protocol, and measured its performance on a range of inputs.

To carry out our experiments, we utilized two standard desktop computers equipped with high-performance 12th Gen Intel(R) Core(TM) i5-12400 2.50GHz processors and 16GB RAM. These computers were connected via a local area network (LAN) with a bandwidth capacity of 100 Mbps. All the experiments were done using two standard desktop computers equipped with 12th Gen Intel(R) Core(TM) i5-12400 2.50GHz processors and 16GB RAM. These computers were connected via a local area network (LAN) with a bandwidth capacity of 100 Mbps. In our experiments, all the elements were randomly generated from some fixed universe, and each party's input set was guaranteed to have no duplicate elements. The time taken by the protocol includes both the execution of oblivious transfer (OT) and the execution phase of garbled circuit. 

\subsection{Plain mSCS}
Table 1 presents the results of a experimental evaluation conducted to assess the performance of the plain mSCS protocol in computing the intersection of large-scale input sets. Our findings reveal that the plain mSCS protocol incurs a significantly higher communication overhead compared to a custom multi-party private set intersection protocol, such as the one proposed by \cite{kolesnikov2017practical}. Specifically, the plain mSCS protocol necessitates between 100 and 1000 times more communication than \cite{kolesnikov2017practical}, which discloses the intersection in plaintext to the participating parties. Despite the increased communication overhead, the results obtained highlight the feasibility of employing the plain mSCS protocol for privacy-preserving set intersection in large-scale scenarios. Furthermore, the protocol demonstrates potential utility in non-real-time applications where private set intersection serves as a submodule.

\subsection{Hashing-mSCS}
Table 2 presents a comprehensive analysis of the minimum numbers of non-free gates required for each element in the Hashing-mSCS protocol, with $\delta$ and $b$ values set at $m=3$ and $\gamma=40$. The number of non-free gates serves as a crucial metric for evaluating the protocol's complexity, as it significantly influences the performance of circuit-based Multi-Party Computation protocols. Notably, the number of non-free gates remains independent of the specific implementation details of the MPC framework, distinguishing it from other benchmarks such as communication overhead or runtime.

By conducting a meticulous analysis and comparing the theoretical computational results of our proposed Hashing-mSCS protocol with the empirical findings from the naive mSCS protocol, we have discovered a substantial reduction of approximately 20\% in communication overhead when employing the Hashing-mSCS protocol. This reduction can be attributed to the utilization of hashing techniques, which partition the elements into distinct bins. Importantly, each bin operates independently, and the intersection of computation results within each bin forms a subset of the final intersection. As a consequence, the Hashing-mSCS protocol facilitates parallel computation, effectively mitigating the challenge of excessive memory consumption associated with large-scale circuits.

This improvement in communication overhead underscores the advantages offered by the Hashing-mSCS protocol in terms of efficiency and scalability. The parallel computation capability, achieved through bin-based partitioning and independent processing, enables more efficient resource utilization. By avoiding the need for storing and processing the entire intersection in a single circuit, the protocol alleviates the burden on memory resources, making it particularly well-suited for scenarios involving large circuit scales. These findings contribute to the growing body of knowledge in the field of secure multi-party computation, paving the way for enhanced privacy-preserving protocols in large-scale settings.

\begin{table}[htbp]
  \centering
    \begin{tabular}{|c|c|c|c|c|}
    \hline
    $logn$ & $\sigma$ & $b$ & $\delta$ & non-free gates \\
    \hline
    8 & 12 & 292 & 0.82 & 1470 \\
    \hline
    8 & 16 & 292 & 0.72 & 1932 \\
    \hline
    8 & 20 & 292 & 0.67 & 2387 \\
    \hline
    12 & 16 & 316 & 0.81 & 1514 \\
    \hline
    12 & 20 & 316 & 0.71 & 1983 \\
    \hline
    12 & 24 & 316 & 0.66 & 2447 \\
    \hline
    16 & 20 & 341 & 0.80 & 1554 \\
    \hline
    16 & 24 & 341 & 0.71 & 2032 \\
    \hline
    16 & 28 & 341 & 0.65 & 2503 \\
    \hline
    24 & 28 & 389 & 0.78 & 1629 \\
    \hline
    24 & 32 & 389 & 0.69 & 2121 \\
    \hline
    32 & 36 & 438 & 0.77 & 1697 \\
    \hline
    32 & 40 & 438 & 0.68 & 2201 \\
    \hline
    \end{tabular}
    \caption{Minimum numbers of non-free gates required for each element in the hashing-mSCS protocol}
\end{table}



%

\bibliographystyle{IEEEtranS}
\bibliography{IEEEabrv,ref}

\begin{thebibliography}{10}
\providecommand{\url}[1]{#1}
\csname url@samestyle\endcsname
\providecommand{\newblock}{\relax}
\providecommand{\bibinfo}[2]{#2}
\providecommand{\BIBentrySTDinterwordspacing}{\spaceskip=0pt\relax}
\providecommand{\BIBentryALTinterwordstretchfactor}{4}
\providecommand{\BIBentryALTinterwordspacing}{\spaceskip=\fontdimen2\font plus
\BIBentryALTinterwordstretchfactor\fontdimen3\font minus
  \fontdimen4\font\relax}
\providecommand{\BIBforeignlanguage}[2]{{%
\expandafter\ifx\csname l@#1\endcsname\relax
\typeout{** WARNING: IEEEtranS.bst: No hyphenation pattern has been}%
\typeout{** loaded for the language `#1'. Using the pattern for}%
\typeout{** the default language instead.}%
\else
\language=\csname l@#1\endcsname
\fi
#2}}
\providecommand{\BIBdecl}{\relax}
\BIBdecl

\bibitem{arbitman2010backyard}
Y.~Arbitman, M.~Naor, and G.~Segev, ``Backyard cuckoo hashing: Constant
  worst-case operations with a succinct representation,'' in \emph{2010 IEEE
  51st Annual symposium on foundations of computer science}.\hskip 1em plus
  0.5em minus 0.4em\relax IEEE, 2010, pp. 787--796.

\bibitem{asharov2013more}
G.~Asharov, Y.~Lindell, T.~Schneider, and M.~Zohner, ``More efficient oblivious
  transfer and extensions for faster secure computation,'' in \emph{Proceedings
  of the 2013 ACM SIGSAC conference on Computer \& communications security},
  2013, pp. 535--548.

\bibitem{blakley1979safeguarding}
G.~R. Blakley, ``Safeguarding cryptographic keys,'' in \emph{Managing
  Requirements Knowledge, International Workshop on}.\hskip 1em plus 0.5em
  minus 0.4em\relax IEEE Computer Society, 1979, pp. 313--313.

\bibitem{cheon2012multi}
J.~H. Cheon, S.~Jarecki, and J.~H. Seo, ``Multi-party privacy-preserving set
  intersection with quasi-linear complexity,'' \emph{IEICE Transactions on
  Fundamentals of Electronics, Communications and Computer Sciences}, vol.~95,
  no.~8, pp. 1366--1378, 2012.

\bibitem{freedman2016efficient}
M.~J. Freedman, C.~Hazay, K.~Nissim, and B.~Pinkas, ``Efficient set
  intersection with simulation-based security,'' \emph{Journal of Cryptology},
  vol.~29, no.~1, pp. 115--155, 2016.

\bibitem{freedman2004efficient}
M.~J. Freedman, K.~Nissim, and B.~Pinkas, ``Efficient private matching and set
  intersection,'' in \emph{Advances in Cryptology-EUROCRYPT 2004: International
  Conference on the Theory and Applications of Cryptographic Techniques,
  Interlaken, Switzerland, May 2-6, 2004. Proceedings 23}.\hskip 1em plus 0.5em
  minus 0.4em\relax Springer, 2004, pp. 1--19.

\bibitem{gao1999k}
Q.~Gao, Y.~Hu, and Z.~Liu, ``K-bitonic sort,'' \emph{Science in China Series E:
  Technological Sciences}, vol.~42, no.~2, pp. 157--164, 1999.

\bibitem{goldwasser1987play}
S.~Goldwasser, ``How to play any mental game, or a completeness theorem for
  protocols with an honest majority,'' \emph{Proc. the Nineteenth Annual ACM
  STOC'87}, pp. 218--229, 1987.

\bibitem{huang2012private}
Y.~Huang, D.~Evans, and J.~Katz, ``Private set intersection: Are garbled
  circuits better than custom protocols?'' in \emph{NDSS}, 2012.

\bibitem{kissner2005privacy}
L.~Kissner and D.~Song, ``Privacy-preserving set operations,'' in
  \emph{Advances in Cryptology--CRYPTO 2005: 25th Annual International
  Cryptology Conference, Santa Barbara, California, USA, August 14-18, 2005.
  Proceedings 25}.\hskip 1em plus 0.5em minus 0.4em\relax Springer, 2005, pp.
  241--257.

\bibitem{kolesnikov2017practical}
V.~Kolesnikov, N.~Matania, B.~Pinkas, M.~Rosulek, and N.~Trieu, ``Practical
  multi-party private set intersection from symmetric-key techniques,'' in
  \emph{Proceedings of the 2017 ACM SIGSAC Conference on Computer and
  Communications Security}, 2017, pp. 1257--1272.

\bibitem{kolesnikov2008improved}
V.~Kolesnikov and T.~Schneider, ``Improved garbled circuit: Free xor gates and
  applications,'' in \emph{Automata, Languages and Programming: 35th
  International Colloquium, ICALP 2008, Reykjavik, Iceland, July 7-11, 2008,
  Proceedings, Part II 35}.\hskip 1em plus 0.5em minus 0.4em\relax Springer,
  2008, pp. 486--498.

\bibitem{pinkas2015phasing}
B.~Pinkas, T.~Schneider, G.~Segev, and M.~Zohner, ``Phasing: Private set
  intersection using permutation-based hashing,'' in \emph{24th $\{$USENIX$\}$
  Security Symposium ($\{$USENIX$\}$ Security 15)}, 2015, pp. 515--530.

\bibitem{pinkas2019efficient}
B.~Pinkas, T.~Schneider, O.~Tkachenko, and A.~Yanai, ``Efficient circuit-based
  psi with linear communication,'' in \emph{Advances in Cryptology--EUROCRYPT
  2019: 38th Annual International Conference on the Theory and Applications of
  Cryptographic Techniques, Darmstadt, Germany, May 19--23, 2019, Proceedings,
  Part III 38}.\hskip 1em plus 0.5em minus 0.4em\relax Springer, 2019, pp.
  122--153.

\bibitem{pinkas2014faster}
B.~Pinkas, T.~Schneider, and M.~Zohner, ``Faster private set intersection based
  on $\{$OT$\}$ extension,'' in \emph{23rd $\{$USENIX$\}$ Security Symposium
  ($\{$USENIX$\}$ Security 14)}, 2014, pp. 797--812.

\bibitem{raab1998balls}
M.~Raab and A.~Steger, ``Balls into bins-a simple and tight analysis,''
  \emph{Randomization and Approximation Techniques in Computer Science}, vol.
  1518, pp. 159--170, 1998.

\bibitem{sang2008privacy}
Y.~Sang and H.~Shen, ``Privacy preserving set intersection based on bilinear
  groups,'' in \emph{Proceedings of the thirty-first Australasian conference on
  Computer science-Volume 74}.\hskip 1em plus 0.5em minus 0.4em\relax Citeseer,
  2008, pp. 47--54.

\bibitem{shamir1979share}
A.~Shamir, ``How to share a secret,'' \emph{Communications of the ACM},
  vol.~22, no.~11, pp. 612--613, 1979.

\bibitem{shamir1980power}
------, ``On the power of commutativity in cryptography,'' in \emph{Automata,
  Languages and Programming: Seventh Colloquium Noordwijkerhout, the
  Netherlands July 14--18, 1980 7}.\hskip 1em plus 0.5em minus 0.4em\relax
  Springer, 1980, pp. 582--595.

\bibitem{yao1986generate}
A.~C.-C. Yao, ``How to generate and exchange secrets,'' in \emph{27th annual
  symposium on foundations of computer science (Sfcs 1986)}.\hskip 1em plus
  0.5em minus 0.4em\relax IEEE, 1986, pp. 162--167.

\bibitem{zahur2015two}
S.~Zahur, M.~Rosulek, and D.~Evans, ``Two halves make a whole: Reducing data
  transfer in garbled circuits using half gates,'' in \emph{Advances in
  Cryptology-EUROCRYPT 2015: 34th Annual International Conference on the Theory
  and Applications of Cryptographic Techniques, Sofia, Bulgaria, April 26-30,
  2015, Proceedings, Part II 34}.\hskip 1em plus 0.5em minus 0.4em\relax
  Springer, 2015, pp. 220--250.

\end{thebibliography}
\cite{huang2012private}
\cite{gao1999k}
\cite{shamir1980power}
\cite{pinkas2015phasing}
\cite{pinkas2019efficient}
\cite{freedman2004efficient}
\cite{kissner2005privacy}
\cite{sang2008privacy}
\cite{cheon2012multi}
\cite{kolesnikov2017practical}
\cite{yao1986generate}
\cite{goldwasser1987play}
\cite{kolesnikov2008improved}
\cite{zahur2015two}
\cite{asharov2013more}
\cite{shamir1979share}
\cite{blakley1979safeguarding}
\cite{pinkas2014faster}
\cite{freedman2016efficient}
\cite{arbitman2010backyard}
\cite{raab1998balls}
\end{document}